\documentclass[twocolumn,showpacs,preprintnumbers,amsmath,amssymb]{revtex4}

\usepackage[dvips]{graphicx}
\usepackage{dcolumn}
\usepackage{bm}

\begin{document}

\preprint{APS/123-QED}

\title{van't Hoff-Arrhenius Analysis of Mesoscopic
and Macroscopic Dynamics of Simple
Biochemical Systems: Stochastic vs.
Nonlinear Bistabilities}

\author{Yunxin Zhang$^1$}
\email{xyz@fudan.edu.cn}
\author{Hao Ge$^1$}%
\email{gehao@fudan.edu.cn}
\author{Hong Qian$^{2,1}$}
\email{hqian@u.washington.edu}
\affiliation{$^1$School of
Mathematical Sciences and Centre for Computational Systems Biology,
Fudan University, Shanghai 200433, PRC.
$^2$Department of Applied Mathematics, University of Washington,
Seattle, WA 98195, USA.
}

\date{\today}

\begin{abstract}
Multistability of mesoscopic, driven biochemical reaction systems
has implications to a wide range of cellular processes. Using
several simple models, we show that one class of bistable chemical
systems has a deterministic counterpart in the nonlinear dynamics
based on the Law of Mass Action, while another class, widely known
as noise-induced stochastic bistability, does not. Observing the
system's volume ($V$) playing a similar role as the inverse
temperature ($\beta$) in classical rate theory, an van't
Hoff-Arrhenius like analysis is introduced.  In one-dimensional
systems, a transition rate between two states, represented in terms
of a barrier in the landscape for the dynamics $\Phi(x,V)$,
$k\propto\exp\{-V\Delta\Phi^{\ddag}(V)\}$, can be understood from a
decomposition $\Delta\Phi^{\ddag}(V)
\approx\Delta\phi_0^{\ddag}+\Delta\phi_1^{\ddag}/V$.  Nonlinear
bistability means $\Delta\phi_0^{\ddag}>0$ while stochastic
bistability has $\Delta\phi_0^{\ddag}<0$ but
$\Delta\phi_1^{\ddag}>0$. Stochastic bistabilities can be viewed as
remants (or ``ghosts'') of nonlinear bifurcations or extinction
phenomenon, and $\Delta\phi_0^{\ddag}$ and $\Delta\phi_1^{\ddag}$ as
``enthalpic'' and ``entropic'' barriers to a transition.
\end{abstract}



\maketitle

    Biochemical reaction dynamics in
a small volume on the order of a cell is stochastic.
For a given spatially homogeneous biochemical kinetic
mechanism, be it a gene regulatory network or an intracellular
signaling pathway, the stochastic trajectories of the
chemical compositions of a mesoscopic, nonlinear reaction
system can be computationally modelled via the Gillespie
algorithm, and its probability distribution
follows the chemical master equation (CME) first
studied by Delbr\"{u}ck \cite{dgp}.
The Delbr\"{u}ck-Gillespie process (DGP) is a multi-dimensional
birth-and-death process \cite{qian_reviews_11}, with the
system's volume, $V$, as a key parameter.
In the limit of infinite large $V$, a macroscopic
nonlinear dynamical system emerges \cite{kurtz}.  This is
precisely the system of ordinary differential equations (ODEs)
following the classic Law of Mass Action (LMA) for chemical
kinetics.

    Bistability in terms of two stable fixed
points is one of the salient features of many
nonlinear chemical reaction systems \cite{schlogl}.
The stationary distribution of a DGP that
corresponds to a macroscopic bistable nonlinear
chemical reaction system is bimodal \cite{vellela_qian_jrsi_09}.
Recently, it also has been discovered that
certain nonlinear reaction system with small $V$
can exhibit bimodal stationary distribution which has
no macroscopic bistable counterpart.  This phenomenon has
been called {\em noise-induced bistability} and
{\em stochastic bifurcation} \cite{NIB,bishop_qian_bj_10}.
Recent experiments in synthetic biological systems
have partially confirmed the theoretical insights
\cite{to_nature_10}.

    One of us has proposed the
notion of {\em nonlinear bistability} and {\em stochastic bistability}
to distinguish the two different
scenarios \cite{qian_reviews_11}.
In particular, it was suggested that there is a
distinct difference in the volume dependence of the
transition rates between states.  While the
transition rate of a state decreases with
increasing $V$ for the nonlinear bistability,
it actually increases for stochastic bistability.
A van't Hoff-Arrhenius-like analysis with respect
to system size $V$ seems possible \cite{hanggi}.

    For a large class of DGP, the stationary
probability distribution with increasing
$V$ has the asymptotic expression \cite{nicolis}
\begin{equation}
        p^{st}_n(V) \approx e^{-V\phi_0(x)-\phi_1(x)}
\label{eq_001}
\end{equation}
where $x=n/V$ is the concentration(s) of the chemical
species, and the $\phi$'s are independent of $V$.
Forthermore, it can be shown that if $\phi_0(x)$
exists and is differentiable, then it is a Lyapunov
function for the macroscopic ODE dynamics
$dx/dt = F(x)$ \cite{hugang}:
\begin{equation}
    \frac{d}{dt}\phi_0(x(t)) =
        \nabla\phi_0(x)\cdot \left(\frac{dx}{dt}
                \right) \le 0.
\end{equation}
Therefore, the stable fixed points of the ODE are
located at the minima of $\phi_0(x)$, and
the leading term in the exponent in Eq. (\ref{eq_001})
indicates that the ``barrier'' between two stable
fixed points increases with $V$.  In the limit of
$V=\infty$, ergodicity breaks down and there
will be no transitions between stable fixed
points (attractors).

    One can making an analogue between $V^{-1}$ and
temperature $T$ in the traditional rate theory \cite{hanggi}. Both
$V^{-1}\rightarrow 0$ and $T$ $\rightarrow 0$ imply a deterministic
limit.  In fact, Arrhenius law states that a rate constant $k\propto
e^{-\Delta G^{\ddag}/k_BT}$, where the activation free energy,
according to van't Hoff analysis, has an enthalpic and an enthropy
part $\Delta G^{\ddag}=\Delta H^{\ddag}-T\Delta S^{\ddag}$.
Therefore negative activation enthalpy leads to decreasing $k$ with
increasing temperature \cite{footnote_1}.

    Comparing the van't Hoff-Arrhenius analysis
with Eq. (\ref{eq_001}), we can identify $\phi_0$
with ``enthalpy'' and $\phi_1$ with entropy.
As we shall show below, stochastic bistability is
associated with a negative $\Delta\phi_0^{\ddag}$.

{\bf\em The general theory|}Let us
consider the 1-dimensional
birth-and-death process with
birth rate $u_n(V)$ and death rate
$w_n(V)$:
\begin{equation}
    \frac{d}{dt}p_n(t) = p_{n-1}u_{n-1}
            -p_n\left(w_n+u_n\right) + p_{n+1}w_{n+1},
\label{bdp}
\end{equation}
$(n\ge 0,u_{-1}=w_0=0.)$
The unique stationary distribution of Eq. (\ref{bdp})
is
\begin{equation}
    p_n^{st} = p_0^{st} \prod_{\ell=0}^{n-1}
        \frac{u_{\ell}(V)}{w_{\ell+1}(V)}
        =p_0^{st} e^{-V\Phi(n,V)},
\end{equation}
where $p_0^{st}$ is a normalization factor, and
$\Phi(n,V)=$
\begin{equation}
     \frac{1}{V}\sum_{\ell=0}^{n-1} \ln\left[
            \frac{w_{\ell+1}(V)}{u_{\ell}(V)}\right]
  = \phi_0(x)+\frac{1}{V}\psi(x,V),
\label{eq_3}
\end{equation}
here we have assumed $n=xV$.  To obtain $\phi_0(x)$, we let $V\rightarrow\infty$ while holding $x$ constant.

    To derive Eq. (\ref{eq_3}), we consider $\ell=Vz$
and noting that functions $u$ and $w$ have
asymptotic expansions according to the
macroscopic LMA
\begin{eqnarray}
     && V^{-1} u_{Vz}(V)
        \approx \mu_{0}(z) + V^{-1} \mu_1(z)
                     + \cdots,
\nonumber\\[-6pt]
\\[-6pt]
    && V^{-1} w_{Vz}(V) \approx \lambda_{0}(z)
                    + V^{-1} \lambda_{1}(z) + \cdots.
\nonumber
\end{eqnarray}
Then we have:
\begin{equation}
    \phi_0(x) = \int_0^x \ln\left(\frac{\lambda_0(z)}
                    {\mu_0(z)}\right)dz.
\end{equation}
We can also obtain a leading order approximation for
$\psi(x,V)\approx \phi_1(x)+V^{-1}\phi_2(x)+\cdots$.  Note that
\begin{eqnarray*}
     && \int_0^x\ln \left(\frac{\lambda_0(z)}
                    {\mu_0(z)}\right)dz
        - \frac{1}{V}\sum_{\ell=0}^{xV-1}\ln\left[
            \frac{\lambda_0(\ell/V)}
            {\mu_0(\ell/V)}\right]
\\
    &=& \sum_{\ell=0}^{xV-1}
        \frac{d}{dz}\left[\ln\frac{\lambda_0(z)}
            {\mu_0(z)}
            \right]_{z=\frac{\ell}{V}}
        \int_{\frac{\ell}{V}}^{\frac{\ell+1}{V}}
        \left(z-\frac{\ell}{V}\right)dz
      + o\left(\frac{1}{V}\right)
\\
        &=& \frac{1}{2V} \int_0^x
        \frac{d}{dz}\left[\ln\frac{\lambda_0(z)}
            {\mu_0(z)}
            \right] dz
             + o\left(V^{-1}\right),
\end{eqnarray*}
and
\begin{eqnarray*}
  && \sum_{\ell=0}^{n-1} \ln\left[\frac{w_{\ell+1}(V)}{u_{\ell}(V)}\right]
-\sum_{\ell=0}^{xV-1}\ln\left[
            \frac{\lambda_0(\ell/V)}
            {\mu_0(\ell/V)}\right]
\\
    &=& \int_0^x \left(\frac{\lambda_1(z)+\lambda'_0(z)}
            {\lambda_0(z)}
            -\frac{\mu_1(z)}{\mu_0(z)}\right)dz
            +o\left(1\right).
\end{eqnarray*}
Therefore,
\begin{eqnarray}
     \phi_1(x) =
         \int_0^x \left(\frac{\lambda_1(z)}{\lambda_0(z)}
                    -\frac{\mu_1(z)}{\mu_0(z)}\right) dz
        +\frac{\ln\left(\mu_0(x)\lambda_0(x)\right)}{2}.
\end{eqnarray}
Therefore, $\Phi(x)\approx \phi_0(x)+\phi_1(x)/V$.

{\bf\em Stochastic bistability|}Stochastic bistability
means for a finite $V$, the $\Phi(x,V)$ has a minimum at
$x=x^*$ and a maximum (or saddle point in multi-dimensional
problems) at $x^{\ddag}$, with $\Phi(x^{\ddag},V)>\Phi(x^*,V)$,
but $\phi_0(x^{\ddag})<\phi_0(x^*)$.  Noting the relation
$\Phi(x,V)\approx\phi_0(x)+V^{-1}\phi_1(x)$, this indicates that
$\phi_1(x^{\ddag})-\phi_1(x^*)=\Delta\phi_1^{\ddag}>0$.

\begin{figure}[t]
\centerline{\includegraphics[width=2.75in,angle=-90]{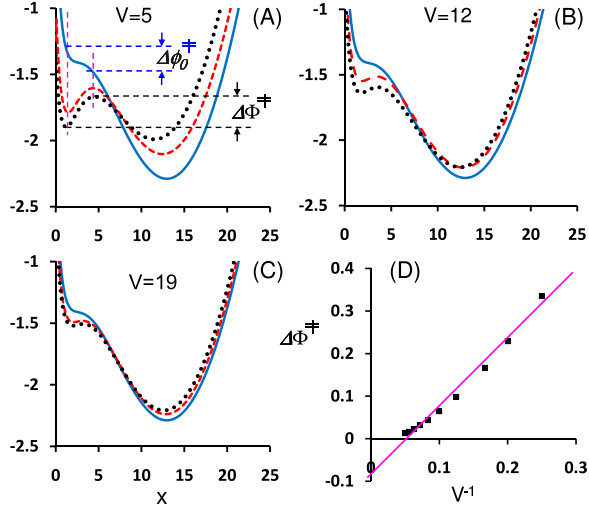}}
\caption{Stochastic bistability has a barrier height
$\Delta\Phi^{\ddag}(V)$ which decreases with increading $V$,
as revealed by the van't Hoff-Arrhenius analysis
$\Delta\Phi^{\ddag}=\Delta\phi_0^{\ddag} +V^{-1}\Delta\phi_1^{\ddag}$.
(A-C): Blue solid line: $\phi_0(x)$
where $x=n/V$ is concentration of $X$ in (\ref{rxn_2});
Black dots: exact $\Phi(x,V)$; Red dashed line: $\phi_0(x)+
V^{-1}\phi_1(x)$ according to Eqs.(\ref{eq_0011},\ref{eq_0012})
with $\theta=0.25,\nu=1.2,\gamma=15$, and different $V$s shown
in the figures.  (D) van't Hoff-Arrhenius
plot, with filled squares for $\Delta\Phi^{\ddag}(V)$, shows a
slope $\sim \Delta\phi_1^{\ddag}>0$ and an interaction
$\sim\Delta\phi_0^{\ddag}<0$.
}
\label{fig_1}
\end{figure}

    We illustrate the theory by an example.  The
Schl\"{o}gl model of nonlinear chemical reactions with
autocatalysis,
\begin{equation}
  A+2X \mathop{\rightleftharpoons}^{\alpha_1}_{\alpha_2}
                3X,  \ \
                X \mathop{\rightleftharpoons}^{\beta_1}_{\beta_2} B,
\label{rxn_2}
\end{equation}
has recently found wide applications in cellular biochemistry
\cite{qian_reviews_11}.  With appropriate parameters it is
well-known to exhibit bistability.  We now consider this
model outside but near its bistable regime.  The stochastic
DGP has birth and death rates \cite{vellela_qian_jrsi_09}:
\begin{eqnarray}
     u_n(V) &=& \frac{k_1n(n-1)}{V}+k_{-2}V,
\nonumber\\[-7pt]
\\[-7pt]
    w_n(V) &=& \frac{k_{-1}n(n-1)(n-2)}{V^2}+k_2n,
\nonumber
\end{eqnarray}
in which $k_1=\alpha_1[A]$, $k_{-1}=\alpha_2$, $k_2=\beta_1$
and $k_{-2}=\beta_2[B]$.  Then,
$\mu_0(z)=k_1z^2+k_{-2}$,
$\mu_1(z)= -k_1z$,
$\lambda_0(z)=k_{-1}z^3+k_2z$,
$\lambda_1(z)=-3k_{-1}z^2$. Then,
\begin{eqnarray}
    \phi_0(x) &=& x\ln\frac{(\theta x^2+\nu\gamma)x}
                 {\nu\gamma(\theta x^2+\nu)}-2\sqrt{\frac{\nu}{\theta}}
                \arctan\left(\sqrt{\frac{\theta}{\nu}}x\right)
\nonumber\\
    &+& 2\sqrt{\frac{\nu\gamma}{\theta}}\arctan\left(
                        \sqrt{\frac{\theta}{\nu\gamma}}x\right)-x,
\label{eq_0011}\\
    \phi_1(x) &=& \frac{1}{2}\ln\left[
            \frac{x(\theta x^2+\nu)^2}
                {(\theta x^2+\nu\gamma)^2}\right],
\label{eq_0012}
\end{eqnarray}
in which
\begin{equation}
        \theta = k_1/k_2, \
         \nu = k_{-2}/k_2, \
          \gamma = k_1k_2/(k_{-1}k_{-2}).
\label{sig_gam}
\end{equation}
Fig. \ref{fig_1} shows that with increasing $V$, the bistability
disappears in the deterministic limit. Furthermore, it shows that
the ``activation enthalpy'' $\Delta\phi_0^{\ddag}$ has a negative
value, and the barrier at finite $V$ is an ``entropic'' one.  This
is the origin of the stochastic bistability.

{\bf\em The ghost of extinction|}A canonical
phosphory\-lation-dephosphorylation signaling with
positive feedbacks exhibits stochastic bistability
\cite{bishop_qian_bj_10}:
\begin{equation}
   E+K+E^* \mathop{\rightleftharpoons}^{\alpha_1}_{\alpha_2}
                2E^*+K,  \ \
                E^*+P \mathop{\rightleftharpoons}^{\beta_1}_{\beta_2} E+P,
\label{rxn_1}
\end{equation}
in which $K$ is a kinase that catalyzes the phosphorylation
reaction $E\rightarrow E^*$, and $P$ is a phosphatase that
catalyzes the dephosphorylation reaction.  The $K$ is only
active, however, after binding an $E^*$ \cite{footnote_2}.
The rates of the DGP are
\begin{equation}
    u_n = \frac{k_1n(N-n)}{V}+k_{-2}(N-n),
    w_n = \frac{k_{-1}n(n-1)}{V}+k_2n,
\end{equation}
where $k_1 =\alpha_1[K]$, $k_{-1}=\alpha_2[K]$,
$k_2=\beta_1[P]$, $k_{-2}=\beta_2[P]$ and $N$
is the total copy number of $E$ and $E^*$ molecules together.
We have
$\mu_0(z)=(k_1z+k_{-2})(e_t-z)$,
$\mu_1(z) = 0$,
$\lambda_0(z) = k_{-1}z^2+k_2z$, and
$\lambda_1(z) = -k_{-1}z$, where $e_t=N/V$.  Therefore,
\begin{eqnarray}
    \phi_0(x) &=& \frac{\theta x
                        +\nu\gamma}{\theta}
                        \ln\left(
                        \frac{\theta}{\nu\gamma}
                         x+1\right)
            +x\ln x
\label{eq_0016}\\
    &-&\frac{\theta x+\nu}{\theta}
        \ln(\theta x+\nu)+(e_t-x)\ln(e_t-x),
\nonumber
\\
    \phi_1(x) &=& \frac{1}{2}\ln\left[\frac{x
            (\theta x+\nu)(e_t-x)}
            {\theta x+\nu\gamma}\right],
\end{eqnarray}
in which
$\theta,\nu,\gamma$ are again given in Eq.
(\ref{sig_gam}).

\begin{figure}[t]
\centerline{\includegraphics[width=1.2in,angle=-90]{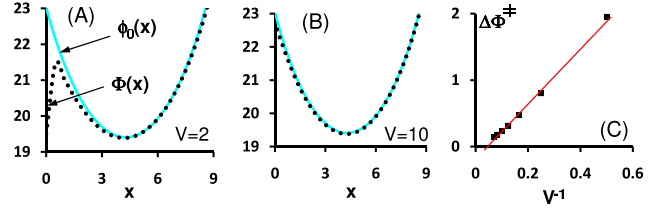}}
\caption{Solid blue lines show $\phi_0(x)$ according to
Eq. (\ref{eq_0016}).  Black dots are $\Phi(x,V)$ following
Eq. (\ref{eq_3}).  Parameters used are
$\theta=0.7,\nu=0.001,\gamma=1000,e_t=10$, with (A) $V=2$
and (B) $V=10$. $x=0$ is a stochatic stable state which disappears
in the deterministic limit. (C) van't Hoff-Arrhenius
plot with filled squares for $\Delta\Phi^{\ddag}$
showing a negative $ \Delta\phi_0^{\ddag}$, as
expected for stochastic bistability.
}
\label{fig_2}
\end{figure}

The stochastic bistability in (\ref{rxn_2}) and Fig. \ref{fig_1}
occurs when the system is near its deterministic
bistable regime. Therefore, one can consider the
stochastic bistability as a ``ghost'' of the saddle-node
bifurcation \cite{strogatz}.  In the present case,
$\phi_0(x)$ in Eq. (\ref{eq_0016}) is a convex function
with a single minimum on $[0,e_t]$ for all parameters.
Therefore, the deterministic dynamics of (\ref{rxn_1})
can not have bistability or bifurcation. It
however, can have a unstable fixed point at $x=0$
when $\beta_2=0$.  This is the case of ``extinction''.
Even though the $x=0$ is unstable to the ODE, the
stationary distribution for the CME is $p_n^{st}=\delta_{n0}$:
The stochastic dynamics goes extinct with probability 1.
As a ``ghost'' of the extinction, this system can also
exhibit stochastic bistability if $\beta_2$ is nonzero
but sufficiently small \cite{bishop_qian_bj_10}.
Fig. \ref{fig_2} shows that while $\phi_0(x)$ is
convex.  However, for finite $V$ there is a
{\em stochastic stable state} at $x=0$.

{\bf\em System with single molecules and stochastic bistability|}We
have so far assumed that each term in the rates $u_n(V)$ and
$w_n(V)$ corresponds to a term in $\mu_0(x)$ and $\lambda_0(x)$.
However, if a chemical reaction system at finite volume contains a
term $O(1)$, then in the limit of $V^{-1}u_{Vx}(V)\sim
\mu_0(x)+V^{-1}\mu_1(x)+\cdots$, the term only contributes to
$\mu_1(x)$.  This is in fact the so-called ``single-molecule
effect''. \cite{wolynes}

{\bf\em DGP with Lewis' chemical detailed balance|} For chemical
system with a single dynamical species, the DGP predicts that
chemical equilibrium has either a binomial distribution (canonical
ensemble) or Poisson distribution (grand canonical ensemble)
following G.N. Lewis' principle of chemical detailed balance
\cite{lewis}.

The canonical ensemble has $u_n=k_+(N-n),w_n =k_-n$
($n\le N$), $\phi_0(x)=x\ln(k_-x/k_+)+(e_t-x)\ln(e_t-x)$, and
$\phi_1(x)=\frac{1}{2}\ln(x(e_t-x))$, with $x\le e_t=N/V$.
$\phi_0(x)$ is convex with a minimum at
$\hat{x}=k_+e_t/(k_++k_-)$.  More importantly,
$V\phi_0(x)+\phi_1(x)$ is convex for $x\in\left(\frac{1}{2V},
e_t-\frac{1}{2V}\right)$.  Note that the asymptotic expansion
$\psi(x,V)=\phi_1(x)+O\left(V^{-1}\right)$ is not
uniformly valid at $x=0,e_t$.  The existence of the
ghost of extinction can not be determined by $\phi_1(x)$.

The grand canonical ensemble has
$u_n=jV,w_n =kn$, $\phi_0(x)=x\ln(x/x^*)-x$,
and $\phi_1(x)=\frac{1}{2}\ln x$, where $x^*=j/k$.
Again, convex $\phi_0(x)$ has a minimum at
$x^*$, and $\phi_1(x)$ is concave. Still
$V\phi_0(x)+\phi_1(x)$ is convex when
$\left(\frac{1}{2V},\infty\right)$.

    Therefore, with detailed balance in a chemical
reaction system, the equilibrium distribution is always unimodal.
The $\gamma$ parameter in the previous sections represents the
energy dissipation of open chemical systems. One can verify that
when $\gamma=1$, the results in the previous two sections are
reduced to what we have here.

{\bf\em  Exit rate of a stochastic stable state|}So far we have
exclusively discussed the stationary distribution in the form of
$e^{-V\Phi(x,V)}$ and its relation to the stochastic bistability.
We now establish the relation between the transition rate from one
stable state to another and the function $\Phi(x,V)$.  In a 1d
birth-and-death process, the  mean time of the first arrival at
$n^*_2$ starting at $n^*_1$, with reflection at $0$
($0<n_1^*<n_2^*$), has been widely used in the various lattice
hopping models \cite{vanKampen}:
\begin{equation}
 T_{n_1^*\rightarrow n_2^*} =
    \sum_{m=n_1^*+1}^{n_2^*}\sum_{n=0}^{m-1}\frac{p^{st}_n(V)}
                {w_m(V)p^{st}_m(V)}
\label{T12_exact}
\end{equation}
in which $p^{st}_n$ is the stationary distribution to
Eq. (\ref{bdp}).  Let $n_1^*=Vx_1^*,n_2^*=Vx_2^*$, then
we found the asymptotic expansion formula
$T_{x_1^*\rightarrow x_2^*}=$
\begin{equation}
     V\int_{x_1^*}^{x_2^*} \frac{1}{\Lambda(x)}
        e^{V\widetilde{\Phi}(x,V)} dx
          \int_0^x e^{-V\widetilde{\Phi}(z,V)}dz
        \left[1+O\left(\frac{1}{V}\right)\right],
\label{d_to_c}
\end{equation}
in which, the modified $\widetilde{\Phi}(x,V)=$
\begin{equation}
        \phi_0(x)+\frac{1}{V}\left\{\phi_1(x)+\ln\frac{\mu_0(x)/\lambda_0(x)-1}{
    \ln(\mu_0(x)/\lambda_0(x))}
            \right\},
\end{equation}
and $\Lambda(x)=$
\begin{equation}
  \frac{1}{\mu_0(x)}
  \left(\frac{\lambda_0(x)-\mu_0(x)}{\ln\lambda_0(x)-\ln\mu_0(x)}
            \right)^2.
\end{equation}
Noting $\Lambda(x)$ playing the role of diffusion coefficient.
Eq. (\ref{d_to_c}) is the same as Kramers' theory for the rate
of crossing a continous energy barrier located at
$x^{\ddag}\in(x_1^*,x_2^*)$ \cite{vanKampen}. Applying Laplace's
method leads
\begin{equation}
    T_{x_1^*\rightarrow x_2^*} \approx
  \frac{2\pi e^{V[\Phi(x^{\ddag},V)-\Phi(x^*_1,V)]}}{\lambda_0(x^{\ddag})
  \sqrt{\phi_0^{''}(x_1^*)|\phi_0^{''}(x^{\ddag})|}}.
\label{T12}
\end{equation}
This simple relation between transition rate and $\Phi(x,V)$ are
unique for 1-d system.  While the concept we developed here will be
valid in higher dimensional systems with saddle point, computation
will be demanding \cite{hanggi}.

{\bf\em Summary|}Nonlinear biochemical reaction systems
can have bi- or multi-stable behavior, which has implications
to a wide range of biological processes such as
epigenetic inheritance, cell differentiation, and
cancer oncogenesis \cite{huang,qian_reviews_11}.
The nature and the existence of the multiple states
are in the nonlinear biochemical reaction schemes and
rates.  For a traditional nonlinear bistable system
in a small volume, the transition rates decreases with
increasing system size.  In the deterministic limit, these
rates become zero.  However, small systems can also
exhibit bistable phenomenon which has no deterministic
counter part. In the latter case, the bistability is
due to the presence of ``noise'', ``stochasticity'', or
``entropic barrier''.  This class of bistable cellular
systems can be quantitatively characterized by an
van't Hoff-Arrhenius like analysis on the volume
dependence of the transition rate(s).  With increasing
volume, the barrier diminishes.  Mathematically, the
transition rate is related to the landscape of the
stochastic dynamics, $k\propto e^{-V\Phi(x,V)}$ where
$\Phi(x,V)$ can be decomposed into $\phi_0(x)+\phi_1/V$.
Nonlinear bistable system has barrier $\Delta\phi_0^{\ddag}>0$
while stochastic bistability has $\Delta\phi_0^{\ddag}<0$
but $\Delta\phi_1^{\ddag}>0$.

\end{document}